\documentclass[12pt,preprint]{aastex}

\slugcomment{To appear in  AJ 142:1 (July 2011)}

\shorttitle{P/2010 A2}
\shortauthors{Jewitt, Stuart, Li}

\begin{document}

%\title{Lost in Plain Sight: Disrupting Asteroid P/2010 A2 \\
%Before Its Discovery}
\title{Pre-Discovery Observations of Disrupting Asteroid P/2010 A2 \\}

\author{David Jewitt$^{1,2,3}$, Joseph S. Stuart$^4$, Jing Li$^2$,  }
\affil{$1$ Dept. Earth and Space Sciences, UCLA \\
$2$ Institute for Geophysics and Planetary Physics, UCLA \\
$3$ Dept. Physics and Astronomy, UCLA \\
$4$ MIT Lincoln Laboratory,
244 Wood Street,
Lexington, MA 02420-9108
}
\email{jewitt@ucla.edu, stuart@ll.mit.edu, jli@igpp.ucla.edu}

\begin{abstract} 
Solar system object P/2010 A2 is the first-noticed example of the aftermath of a recently disrupted asteroid, probably resulting from a collision.   Nearly a year elapsed between its inferred initiation in early 2009 and its eventual detection in early 2010.  Here, we use new observations to assess the factors underlying the visibility, especially to understand the delayed discovery. We present prediscovery observations from the LINEAR telescope and set limits to the early-time brightness from SOHO and STEREO satellite coronagraphic images.  Consideration of the circumstances of discovery of P/2010 A2 suggests that similar objects must be common, and that future all-sky surveys will reveal them in large numbers.
 
\end{abstract}


\section{Introduction}
\begin{figure}
\epsscale{1.0}
\begin{center}
%%\plotfiddle{f101.ps}{0.5cm}{-90.}{.4}{.4}{-10000}{.00}
%\plotone{figure6.pdf}
%\plotone{discovery.pdf}
\plotone{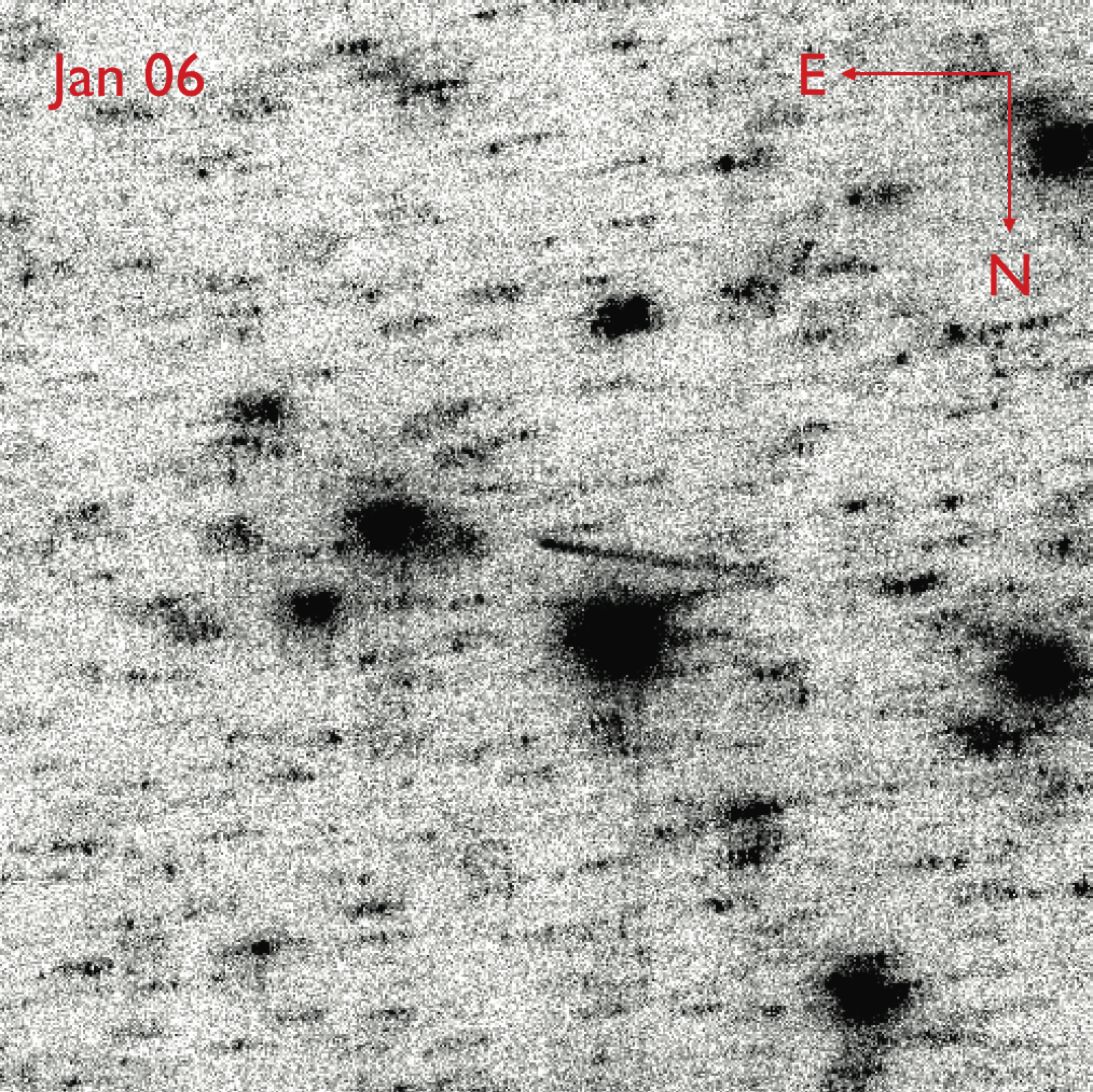}
\caption{Discovery image of P/2010 A2 from UT 2010 January 06 at the LINEAR survey telescope.  This is the median composite of 5 images each shifted to account for the non-sidereal motion of the target.  The region shown is 500$\times$500 pixels (1130$\times$1130$\arcsec$) in size.  Direction arrows are shown.  \label{discovery} } 
\end{center} 
\end{figure}

Solar system object P/2010 A2 was discovered on UT 2010 January 6 (Kadota et al.~2010) using the LINEAR (Lincoln Near-Earth Asteroid Research) survey telescope (Stokes et al.~2000). The discovery image is shown in Figure (\ref{discovery}).   The semimajor axis, eccentricity and inclination are 2.291 AU, 0.124 and 5.3$\degr$, respectively, all typical of an inner-belt asteroid. The Tisserand invariant with respect to Jupiter is $T_J$ = 3.58, showing dynamical decoupling from Jupiter and confirming its asteroid-like dynamical nature.  Ground-based observations quickly showed P/2010 A2 to be an object of special interest, with an appearance unlike those of either asteroids or, in detail, comets (Jewitt et al.~2010a, Licandro et al.~2010).  In particular, whereas in most comets the nucleus is  embedded in a coma of recently-released dust, P/2010 A2 showed a very faint nucleus separated from a narrow, parallel-sided dust structure.  This characteristic suggested impulsive emission followed by a period of relative inactivity in which radiation pressure acted to separate the dust from the nucleus.  Subsequent, high-resolution morphological observations from the Hubble Space Telescope confirmed this basic picture, showing dust released from trailing sources (presumably large boulders) arranged in filamentary structures behind the main nucleus (Jewitt et al.~2010b).  The changing position angle of the tail of P/2010 A2 in the Hubble data is consistent with impulsive formation in 2009 February/March (see also Snodgrass et al.~2010).  (A contradictory result, in support of continuous emission over an 8-month period, was reported by Moreno et al.~(2010).  However, their conclusion is based on the analysis of data from only a 9-day interval in 2010 January, and lacks the benefit afforded by changing observing geometry in data obtained over the following months). 

The impulsive formation in early 2009 predates the discovery by roughly 10 months and begs the question ``why was P/2010 A2 not discovered earlier?''.   In this paper, we will describe observations from orbiting solar observatories that had the potential to record the newly-formed P/2010 A2 using coronagraphic imaging.  We will also describe observations from ground-based telescopes in later months that could have detected P/2010 A2 before its eventual discovery in January 2010.

\section{Observations}
\begin{figure}
\epsscale{1.0}
\begin{center}
%%\plotfiddle{f101.ps}{0.5cm}{-90.}{.4}{.4}{-10000}{.00}
%\plotone{figure6.pdf}
%\plotone{RDa.pdf}
\plotone{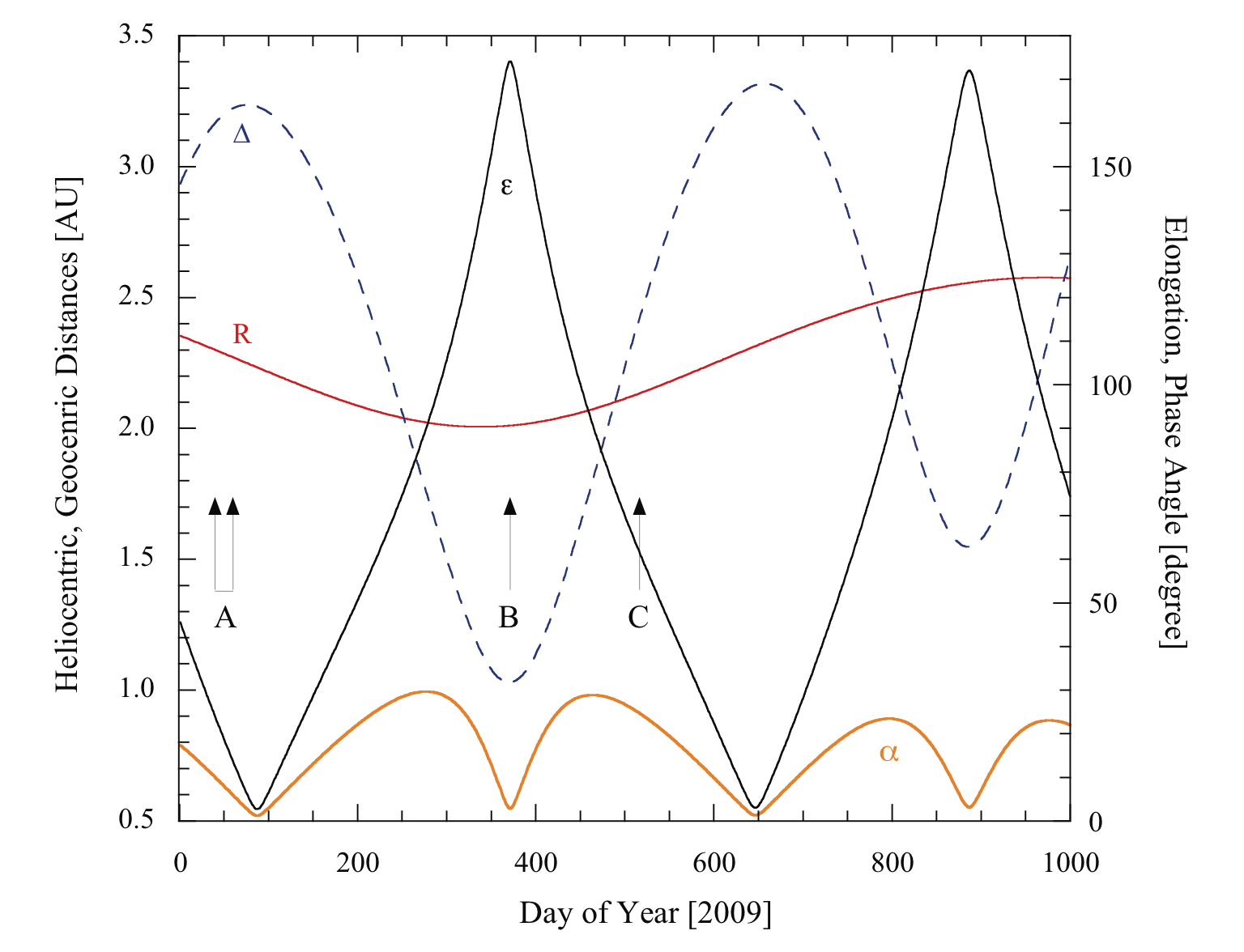}
\caption{Time variation of (left axis) the heliocentric and geocentric distances and (right axis) the elongation and phase angles for a terrestrial observer.  Time is measured in number of days since UT 2009 January 1.  Arrows denote A: estimated time of the disruption from (left arrow A) Snodgrass et al.~2010 and (right arrow A) Jewitt et al.~2010, B: time of the discovery of P/2010 A2 and C: the time of the last Hubble Space Telescope observation.  \label{RDa} } 
\end{center} 
\end{figure}

In Figure (\ref{RDa}) we show the time dependence of the heliocentric, $R$, and geocentric, $\Delta$, distances (both in AU) and of the elongation and phase angles, $\epsilon$ and $\alpha$, respectively  (both in degrees).  All quantities are computed for a geocentric observer.  The time is measured by DOY (Day of Year), defined such that UT 2009 January 1.0 (Julian Day 2454832.5) corresponds to DOY  = 1, UT 2010 January 1 is DOY = 366 and so on.   Plotted for reference on the figure are arrows showing A: two estimates of the time of the disruption of P/2010 A2 (UT 2009 Feb 10 from Snodgrass et al.~2010 and UT 2009 March 1 from Jewitt et al.~2010b), B:  UT 2010 January 6, the date of the discovery (Kadota et al.~2010)  and C: UT 2010 May 29, the date of the last Hubble Space Telescope observation used by Jewitt et al.~(2010b) to characterize the object.  Figure (\ref{RDa}) immediately shows that P/2010 A2 disrupted at a time when its solar elongation was $\epsilon \leq$30$\degr$, placing it in the daytime sky as seen from Earth.  Night-time telescopes typically survey the sky at elongations
 $\epsilon >$ 60$\degr$, a value not reached by P/2010 A2 until approximately DOY 222 (UT 2010 August 10).  This is still five months before the eventual discovery of P/2010 A2 on DOY 371 (UT 2010 January 6).
 
Next, we examine geometrical effects on the brightness of P/2010 A2.  Assuming, for the moment, that P/2010 A2 has the photometric behavior of a solid-body, we may write that the apparent magnitude, $V$, is related to the absolute magnitude, $V(1,1,0)$, (the magnitude to be observed at $R$ = $\Delta$ = 1 AU and $\alpha$ = 0$\degr$) by $V$ = $V(1,1,0) + \Delta V$, where

\begin{equation}
\Delta V = 5 \log_{10} \left(R \Delta \right) + 2.5 \log_{10} \left( \Phi(\alpha) \right).
\label{dmag}
\end{equation}

\noindent This is simply the inverse square law of brightness, with $\Phi(\alpha)$ an additional term to account for phase-angle dependent brightness variations.  For solid bodies, it is common to write the simple form $\Phi(\alpha)$ = 10$^{0.4 \beta \alpha}$, where $\beta$ is the phase coefficient, measured in magnitudes per degree of phase angle.  %Since the range of phase angles swept by P/2010 A2 is modest (0 $\leq \alpha \leq$ 30$\degr$; see Figure (\ref{RDa})), the adopted value of $\beta$ is not particularly important. 
The range of values exhibited by many asteroids and cometary nuclei is 0.02 $\leq \beta \leq$ 0.06.  Phase effects of comparable size have been measured also in active comets (Schleicher et al.~1998) over this narrow range of phase angles.  More complicated forms for the phase term have been proposed for use on asteroids. We have no reason to suppose that these should describe the phase term due to particle scattering in P/2010 A2 any better than the simple form employed, and so we do not use them here.

\begin{figure}
\epsscale{1.0}
\begin{center}
%%\plotfiddle{f101.ps}{0.5cm}{-90.}{.4}{.4}{-10000}{.00}
%\plotone{figure6.pdf}
%\plotone{Dm.pdf}
\plotone{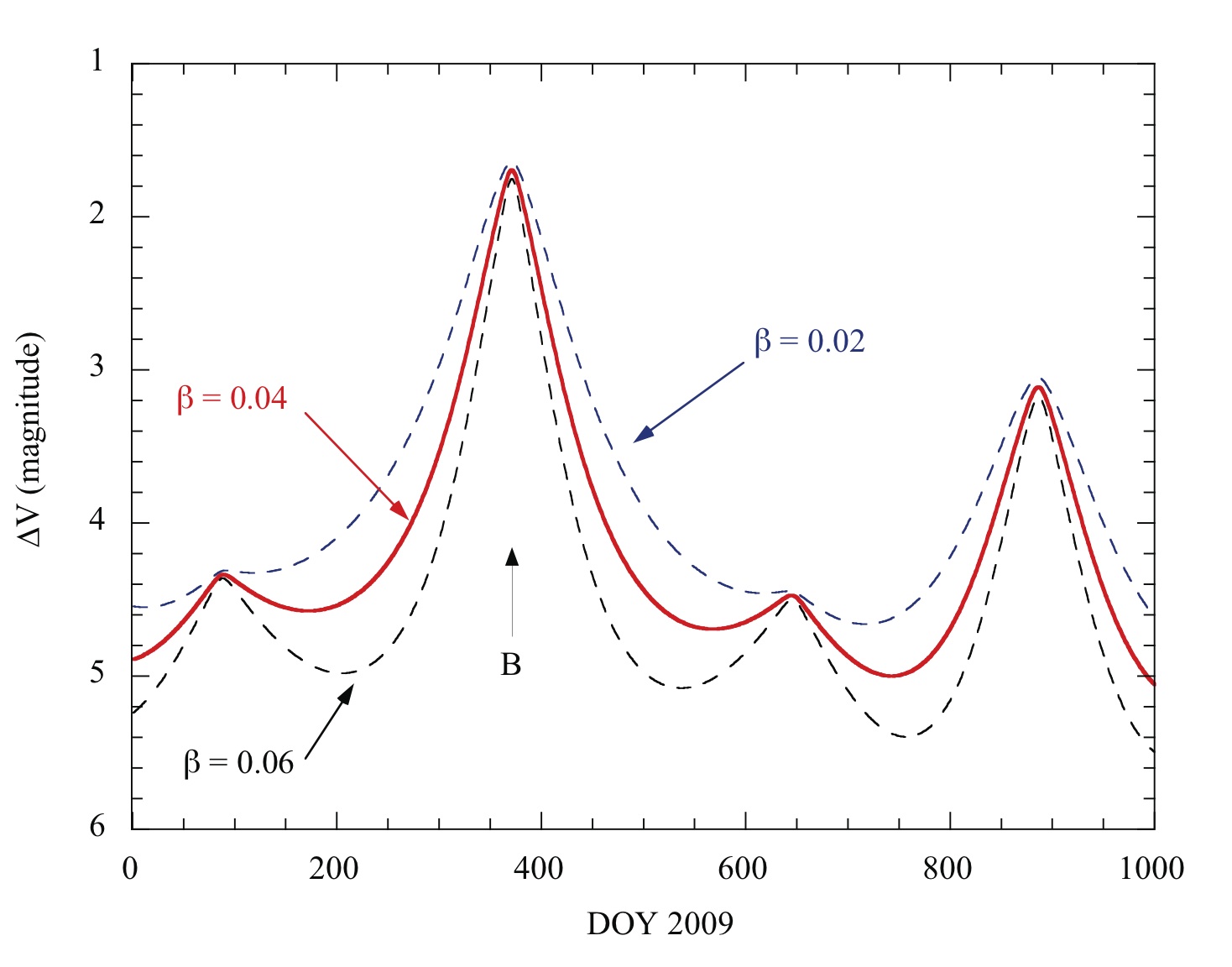}
\caption{Geometric dimming factor (in magnitudes) as a function of time, measured in number of days from UT 2009 January 1.  The date of the discovery of P/2010 A2 is marked.   \label{Dm} } 
\end{center} 
\end{figure}

Figure (\ref{Dm}) shows Equation (\ref{dmag}) for assumed phase function parameters $\beta$ = 0.02, 0.04 and 0.06 magnitudes per degree, for the same range of dates as in Figure (\ref{RDa}).  The figure is plotted with an inverted y-axis to show that, all else being equal, P/2010 A2 should have been brightest near DOY 370, closely matching the date of discovery (marked B in Figure (\ref{Dm}).  The largest influence on $\Delta V$ is the geocentric distance, which reached a minimum near this time, but $R$ and $\alpha$ were also small, further reducing $\Delta V$.  The proximity between the peak in Figure (\ref{Dm}) and the date of discovery is unlikely to be a coincidence. It suggests that the object was not discovered at earlier times, in part, because of geometrical dimming.  For example, on DOY 222, when the object first reached $\epsilon$ = 60$\degr$, the apparent brightness was fainter than at the peak by $\sim$2.5 magnitudes (a factor of 10), according to Figure (\ref{Dm}).

\subsection{LINEAR}

The LINEAR survey obtained discovery and pre-discovery observations of P/2010 A2 using the LINEAR-1 telescope located near Socorro, New Mexico.  The telescope is a GEODSS-type (Ground-based Electro-Optical Deep Space Surveillance) telescope with a $1$-m diameter aperture, and f/2.15 optics. The camera contains a custom-built 2560x1960 pixel frame-transfer charge-coupled imaging device.  The angular pixel scale is 2.25$\arcsec$ pixel$^{-1}$ and the field of view approximately  1.6$\degr$$\times$1.2$\degr$ (Stokes et al.~2000).  Analyses presented here were based on sub-frames of approximately 500$\times$500 pixels (1130$\times$1130$\arcsec$) extracted from the original data. The full-width-at-half maximum of the point-spread function in individual images was $\sim$2.5 pixels, corresponding to $\sim$5.7$\arcsec$. The limiting magnitude of the survey varies depending on observing conditions, and is especially sensitive to the phase and angular distance of the moon.

%Observations that included the expected position of P/2010 A2 are listed in Table (\ref{linear}).  

\begin{figure}
\epsscale{0.9}
\begin{center}
%%\plotfiddle{f101.ps}{0.5cm}{-90.}{.4}{.4}{-10000}{.00}
%\plotone{figure6.pdf}
%\plotone{LINEAR_composite.pdf}
\plotone{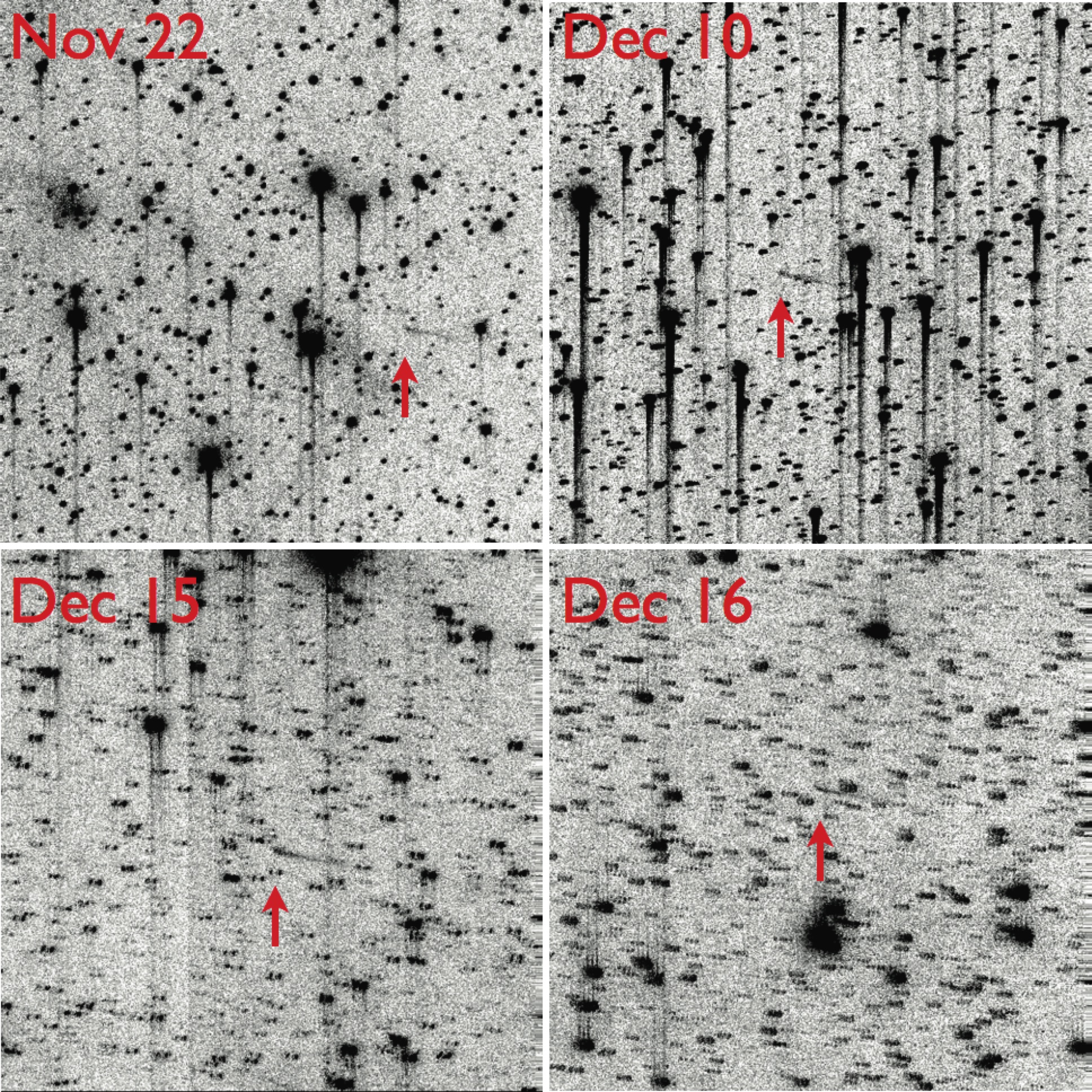}
\caption{Prediscovery images of P/2010 A2 recorded using the LINEAR survey telescope.   These are the median combinations of five consecutive integrations of 9.2-9.8 seconds each, shifted for the motion of the object and taken on UT 2009 Nov 22, Dec 10, 15 and 16, as marked.  Arrows show the location of the object.  Each panel is 500$\times$500 pixels (1130$\times$1130$\arcsec$) in size, and has North to the bottom, East to the left.  The panel sizes, orientations, methods of computation and image stretches are the same as used in Figure \ref{discovery}.  \label{composite} } 
\end{center} 
\end{figure}

P/2010 A2 fell within the field of view of LINEAR on twelve nights in 2009.  We visually searched images from UT 2009 Aug 28, Sep 15, 22 and 28, Oct 15, 23 and 31 and Nov 11 without success.  In many cases, the expected position of the object overlaps the scattered light from bright field stars, severely compromising the effective detection limits.  However, P/2010 A2 is clearly recorded in prediscovery observations from UT 2009 Nov 22, Dec 10, 15 and 16.  Images from these nights, computed from the sum of five integrations each of 9.2-9.8 seconds integration (total 46-49 seconds), are shown in Figure (\ref{composite}).  That P/2010 A2 was missed in these images by the object-finding software is easily understood as a consequence of the low surface brightness and peculiar morphology of the dust tail.  The software hunts for local brightness maxima, using adjacent pixels to define the background, and then searches for linear, correlated motion between bright pixels identified in the five images of each field (Stokes et al.~2000).  In the case of P/2010 A2, field stars, image defects and noise in the data produce false local brightness maxima around and on the dust trail, hiding the object from the software.   

The main practical concern with photometry of a low surface brightness source such as P/2010 A2 lies in the determination of the sky background.  The large pixel size of the LINEAR data further complicates photometry by causing frequent contamination of the faint cometary dust tail by bright field stars, as may be seen in Figures (\ref{discovery}) and (\ref{composite}).  To measure the brightness of the object, we first digitally removed nearby stars by replacing affected pixels with the average of the surrounding pixel values.  Next, we calculated the median of the five consecutive LINEAR images taken each night (for a total integration of 46-49 seconds) and then rotated the image to bring the axis of the dust tail into a horizontal position.  Photometry used an aligned rectangular aperture 59$\times$10 pixels (133.5$\arcsec \times$22.6$\arcsec$), with the sky computed from a contiguous set of rectangular apertures of identical size but displaced perpendicular to the tail axis.  This box size deliberately excludes the distant tail of P/2010 A2 in order to reduce photometric uncertainties caused by the sky background.  The result is that the reported magnitudes slightly underestimate the total brightness, but the effect is small ($\sim$10\%) and of no importance in the following discussion of the lightcurve. Absolute calibration of the LINEAR data was obtained using photometry of field stars having $R$-band magnitudes from the USNO-B star catalogue, with typically 4 stars used per field (Monet et al.~2003).  We converted to $V$-magnitudes assuming $V$-$R$ = 0.5.  The resulting magnitudes are listed in Table (\ref{phot}), together with uncertainties estimated from the scatter in the sky background counts.  Additional, systematic uncertainties caused by the non-standard photometric passband of LINEAR have been ignored.

\subsection{Other Ground-based Telescopes}
Fields observed by the Catalina Sky Survey telescope (a 0.7-m diameter, f/1.9 Schmidt design; Drake et al.~2009) did not include P/2010 A2 on any occasion in the period August - October 2009 (private communication, Ed Beshore, 2010 July 23).   The SpaceWatch survey fields (using a 0.9-m diameter, f/3 telescope) likewise missed P/2010 A2 before November 2009 (private communication, Robert McMillan, 2010 July 15).  Both telescopes routinely detect asteroids at $V \sim$ 19 to 20.  We are unaware of any other pre-discovery observations of P/2010 A2 from ground-based telescopes.

\subsection{SOHO}
Next, we examined coronagraphic images from NASA's SOHO spacecraft.  The LASCO C3 (``clear'' filter) coronagraph has a 1024$\times$1024 pixel CCD camera with 56$\arcsec$ pixels giving a 16$\degr$ wide field of view (Brueckner et al.~1995, Morrill et al.~2006).  The quantum efficiency of the instrument peaks at $\sim$0.3 near wavelength $\lambda$ = 7000\AA, falling to half the peak value at $\lambda \sim$ 5000\AA~in the blue and $\lambda \sim$ 8500\AA~in the red, giving FWHM = 2500\AA. The central 2$\degr$ are obscured by an occulting mask while an additional region is shadowed and vignetted by a pylon supporting this mask.  Images are recorded with integration time 19s.  

\begin{figure}
\epsscale{0.9}
\begin{center}
%%\plotfiddle{f101.ps}{0.5cm}{-90.}{.4}{.4}{-10000}{.00}
%\plotone{figure6.pdf}
%\plotone{soho.pdf}
\plotone{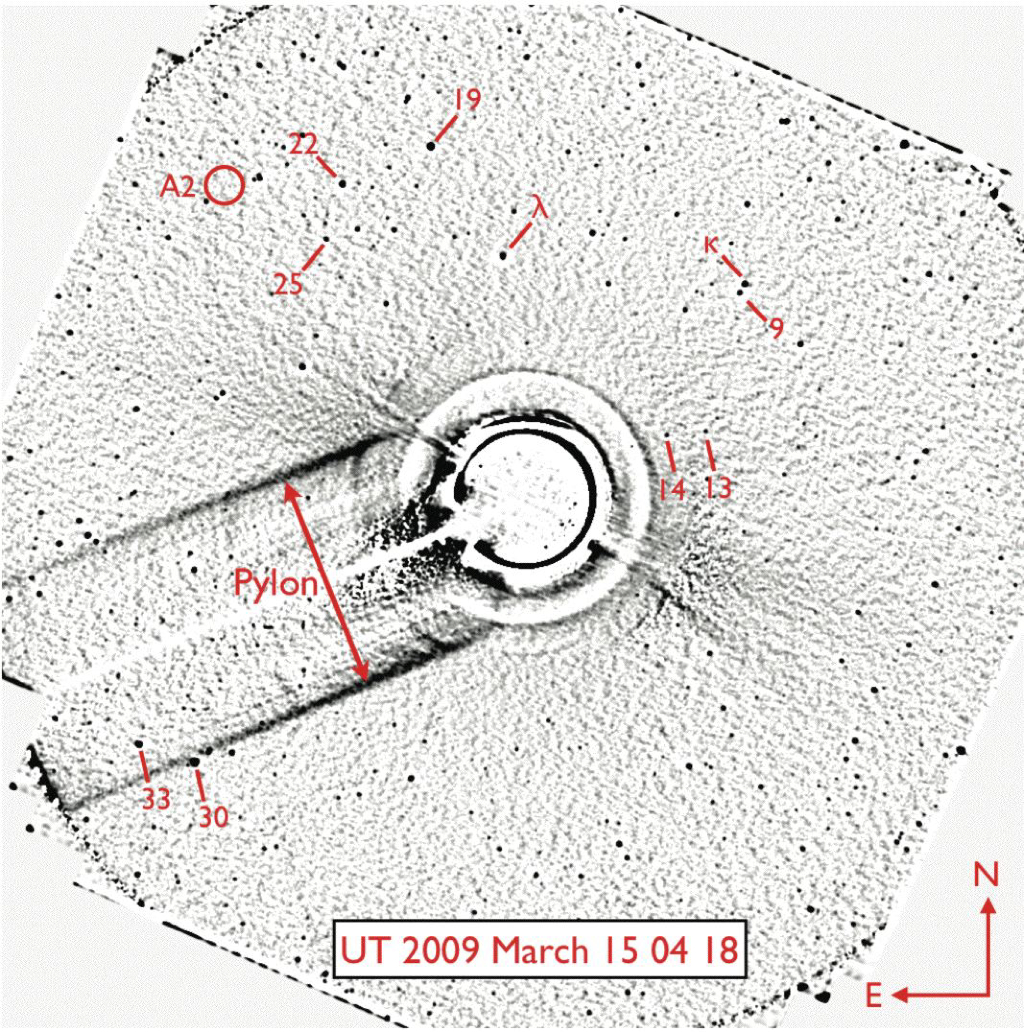}
\caption{Sample LASCO C3 coronographic image from the SOHO satellite, taken UT 2009 March 15d 04h 18m.  A median image has been subtracted to remove large scale brightness gradients.   Labels identify stars in the constellation of Pisces (see text) while the position of P/2010 A2 is circled.  The region affected by obstruction and vignetting from the support pylon is also marked. Stars and other signals appear black in this reverse-polarity image.  The field of view is approximately 16$\degr$ wide.   \label{soho} } 
\end{center} 
\end{figure}

The elongation of P/2010 A2 as seen from SOHO prevented observations before about UT 2009 March 14 and after April 12. We used the earliest high quality LASCO C3 images, from UT 2009 March 15, when the elongation was $\sim$6.8$\degr$ (Figure \ref{soho}).  On this date the target has moved inwards from the edge of the field of view, where vignetting is a concern, but remains far from the Sun, where the sensitivity is limited by the bright coronal background.  To reduce the effects of strong brightness gradients caused by the background solar corona, we subtracted an image computed from the median of LASCO C3 images taken over a surrounding 1 day interval.  To further improve the image, but only for clarity of presentation, we have lightly smoothed the image in Figure \ref{soho}.  Fine structures caused by time-variable components in the corona remain in the image, but strong radial gradients are successfully removed.  We blinked background-subtracted images visually in order  to search for A2, without success.    

The effective limiting magnitude of the LASCO C3 data is a strong function of angular distance from the Sun because of the strongly centrally concentrated solar corona.  For reference, we have identified field stars in Figure \ref{soho} as follows: 13 (5.73), 14 (5.72), 19 (4.98), 22 (4.76), 25 (6.28), $\lambda$ (4.39) where, for example 13 refers to 13 Psc and (5.73) is the $R$ magnitude of this star from the USNO B catalog.  To estimate the maximum possible brightness of P/2010 A2 we measured the instrumental magnitudes of nearby bright stars (19 Psc, 22 Psc and 25 Psc), then scaled these to the faintest objects visible in the vicinity of the expected position of P/2010 A2.   In this way, we set a limit to the apparent brightness of P/2010 A2 of  $V \geq$ 7.9, under the assumption that the object had a point-like morphology on UT 2009 Mar 15.  Because of the nature of the LASCO data, especially the large pixel size, the spatially complex coronal background, and the high density of ``cosmic ray'' events, we believe that the uncertainty on this limiting magnitude approaches a full magnitude.  

\subsection{STEREO}
%STEREO consists of two spacecraft moving approximately in the orbit of the Earth but separating from it slowly in opposite directions (Kaiser et al.~2008).  The two spacecraft house cameras which image the Solar corona simultaneously from different perspectives, and hence provide three-dimensional information.  The spacecraft ahead of Earth (STEREO-A) images the corona to the east of the Sun as seen from its position, while the trailing spacecraft (STEREO-B) images corona to the west of the Sun.  By bad luck, at the predicted time of its outburst, P/21010 A2 appeared to the west of the Sun from STEREO-A, and to the east of the Sun from STEREO-B, and was therefore invisible from both.  

STEREO consists of two spacecraft moving separately in the orbit of the Earth but separating from it slowly in opposite directions  (Kaiser et al.~2008). The two spacecraft house identical instrument packages which observe the solar corona and heliospheric environment simultaneously from different perspectives, and hence provide three-dimensional information. One package is the Sun Earth Connection Coronal and Heliospheric Investigation (SECCHI, Howard et al.~2008) which includes two wide angle visible light Heliospheric Imagers (HI, Eyles et al.~2009).  These have field centers offset from the Sun by $\pm$14.0$\degr$ (HI-1) and $\pm$53.7$\degr$ (HI-2), and fields of view 20$\degr$ and 70$\degr$ wide, respectively. At the inferred time of the A2 outburst, the HI camera on the spacecraft moving ahead of the Earth (STEREO-A) imaged the corona to the east of the Sun as seen from its position, while the HI camera on the trailing spacecraft (STEREO-B) imaged corona to the west of the Sun. By bad luck, at the predicted time of its outburst, P/2010 A2 appeared to the west of the Sun from STEREO-A, and to the east of the Sun from STEREO-B, and was therefore invisible from both.

The SECCHI package also includes two white light coronagraphs. The first, called COR1, has a Sun-centered field of view covering an annulus with inner and outer dimensions of 1.4-4.0 solar radii (0.7$\degr$-2.0$\degr$), while the second, COR2, records an annular region extending from 2.0-15.0 solar radii (1.0$\degr$-7.5$\degr$) (Kaiser et al.~2008). The elongation of P/2010 A2 fell within the 0.7$\degr$-7.5$\degr$ range as seen from STEREO A in between UT 2009 February 01 - 10 while from STEREO B the corresponding dates of visibility were 2009 May 24 - June 24. The predicted date of the impact event occurred in February/March 2009 (Jewitt et al.~2010, Snodgrass et al.~2010).  We examined SECCHI coronagraph images for both time periods, recognizing that the STEREO A observations might have been taken before the event, given the uncertainties in the predicted date. In any case, P/2010 A2 was invisible in both February and June periods down to a limiting magnitude $V \sim$ 8, as estimated from nearby field stars.

%[consider plot for elong vs date - color coded for east vs west (or signed]

\section{Discussion}
Observations of P/2010 A2 in the interval 2010 January - May show that the dust tail is dominated by particles in the millimeter to centimeter size range (Jewitt et al.~2010, Moreno et al.~2010, Snodgrass et al.~2010).  Furthermore, the variation of the surface brightness with position along the tail (caused by size-dependent radiation pressure separation) is consistent with a power law distribution of particle sizes,

\begin{equation}
n(a)da = \Gamma a^{-q} da
\label{power}
\end{equation}

\noindent where $n(a)da$ is the number of particles having radii in the range $a$ to $a + da$ and $\Gamma$ and $q$ are constants of the distribution.  In the radius range 0.1 $\le a \le$ 1 cm, the best-fit size distribution index is $q$ = 3.3$\pm$0.2 (Jewitt et al.~2010).  Independently, Moreno et al.~(2010) derived $q$ = 3.4$\pm$0.3, while Snodgrass et al.~(2010) reported $q \sim$3.5, but did not state an uncertainty.  

It is possible that \textit{only} particles in the $mm$ to $cm$ size range were expelled from P/2010 A2, perhaps representing some fundamental scale of granularity in the material of which the parent body was made. If so, the measured cross-section of the tail in data from 2010 is a good representation of the cross-section at the epoch of ejection, because these large particles have not been removed by the action of radiation pressure.  However, it is more likely that smaller particles, perhaps containing a much larger scattering cross-section, were present at formation but were swept away by radiation pressure in the year before discovery.  In this case, P/2010 A2 would have been intrinsically much brighter near the time of its formation than when observed in 2010.  But by how much?

To estimate the possible initial brightness of P/2010 A2 we proceed as follows.  We assume that the dust particles are distributed in size according to Equation (\ref{power}), over a range extending from a minimum particle radius, $a_{min}$, to a maximum radius, $a_{max}$.  The ratio of the total geometric cross-section to the cross-section of $mm$ to $cm$ sized particles in the power law distribution is given by 

\begin{equation}
f = \frac{\int_{a_{min}}^{a_{max}}\pi a^2 n(a) da }{\int_{0.1}^{1} \pi a^2 n(a) da},
\label{factor}
\end{equation}

\noindent where all particle radii are expressed in centimeters and the integral in the denominator represents the combined  cross-section of the mm to cm-sized grains.  We adopt $a_{min}$ = 0.1 $\mu$m (10$^{-5}$ cm), since the scattering efficiencies of smaller particles are very low at optical wavelengths and much smaller particles will not contribute significantly to the optical signal (Bohren and Huffman 1983).  The value of $a_{max}$ is unknown, except that $a_{max} \geq$ 1 cm as shown by the Hubble Telescope images, and we leave it as a free parameter in our calculations.  

\begin{figure}
\epsscale{0.9}
\begin{center}
%%\plotfiddle{f101.ps}{0.5cm}{-90.}{.4}{.4}{-10000}{.00}
%\plotone{figure6.pdf}
%\plotone{f_plot.pdf}
\plotone{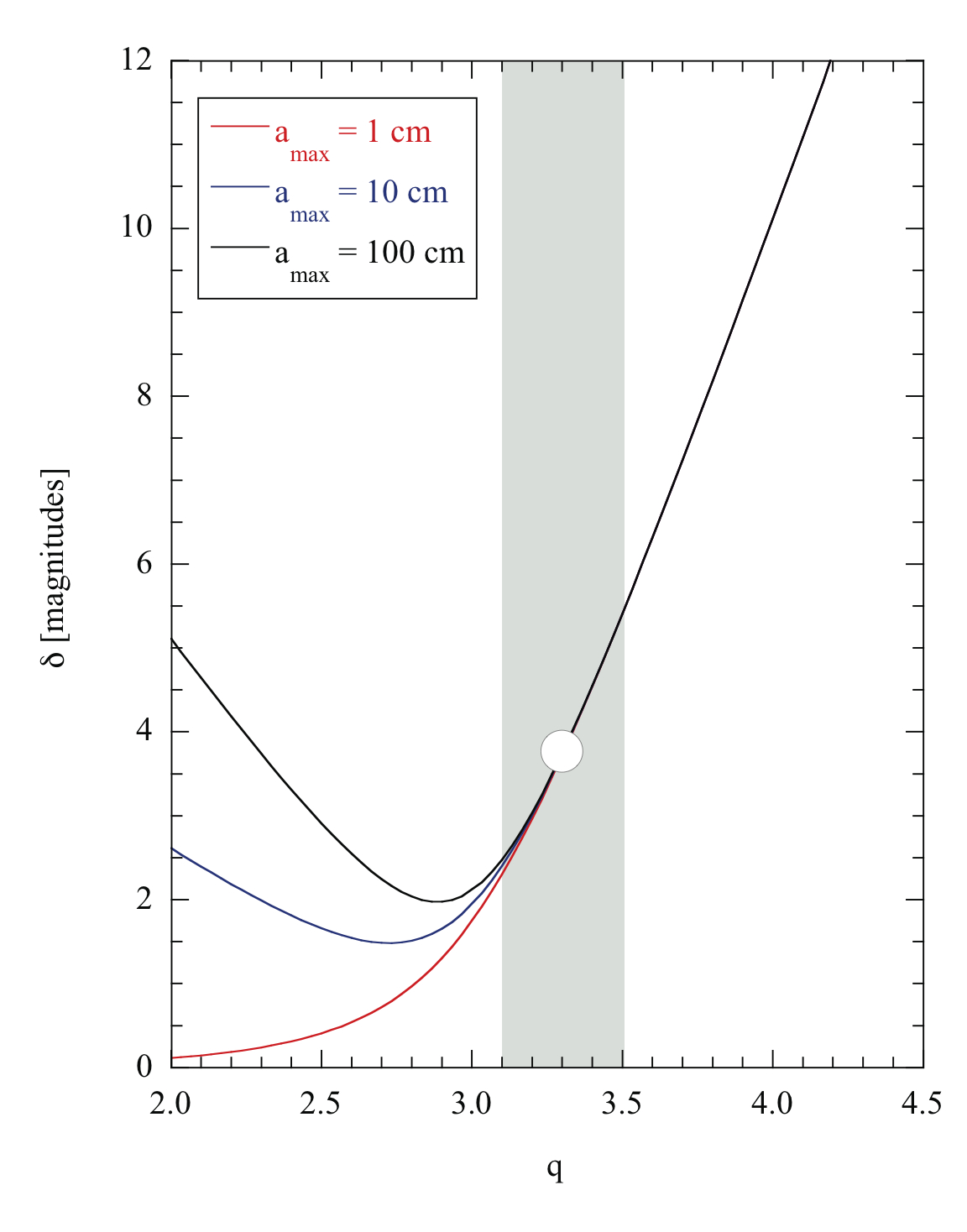}
\caption{Brightness enhancement, $\delta$ = 2.5$\log_{10}f$, from Equation (\ref{factor}).  The three curves show $\delta$ for different values of $a_{max}$.  The circle shows the best-fit value of $q$ from Jewitt et al.~(2010b) while the shaded region marks the formal uncertainty on $q$.  \label{f} } 
\end{center} 
\end{figure}

Figure (\ref{f}) shows the magnitude difference, $\delta$ = 2.5 $\log_{10}f$, computed from Equations (\ref{power}) and (\ref{factor}) and plotted as a function of $q$, for different values of $a_{max}$.  As expected, the value of $a_{max}$ is unimportant for distributions having $q >$ 3, since large particles then carry a small fraction of the total cross-section.  With $q$ = 3.3$\pm$0.2, the Figure shows that $\delta$ lies in the range 2.3 $\le \delta \le$ 5.5 magnitudes.  In other words, if small dust particles were initially present in abundances consistent with the measured size distribution, then the initial absolute magnitude of P/2010 A2 would have been brighter than at discovery by $\delta \sim$2.3 to 5.5 magnitudes.

\begin{figure}
\epsscale{1.0}
\begin{center}
%%\plotfiddle{f101.ps}{0.5cm}{-90.}{.4}{.4}{-10000}{.00}
%\plotone{figure6.pdf}
%\plotone{Photometry_plot_2.pdf}
\plotone{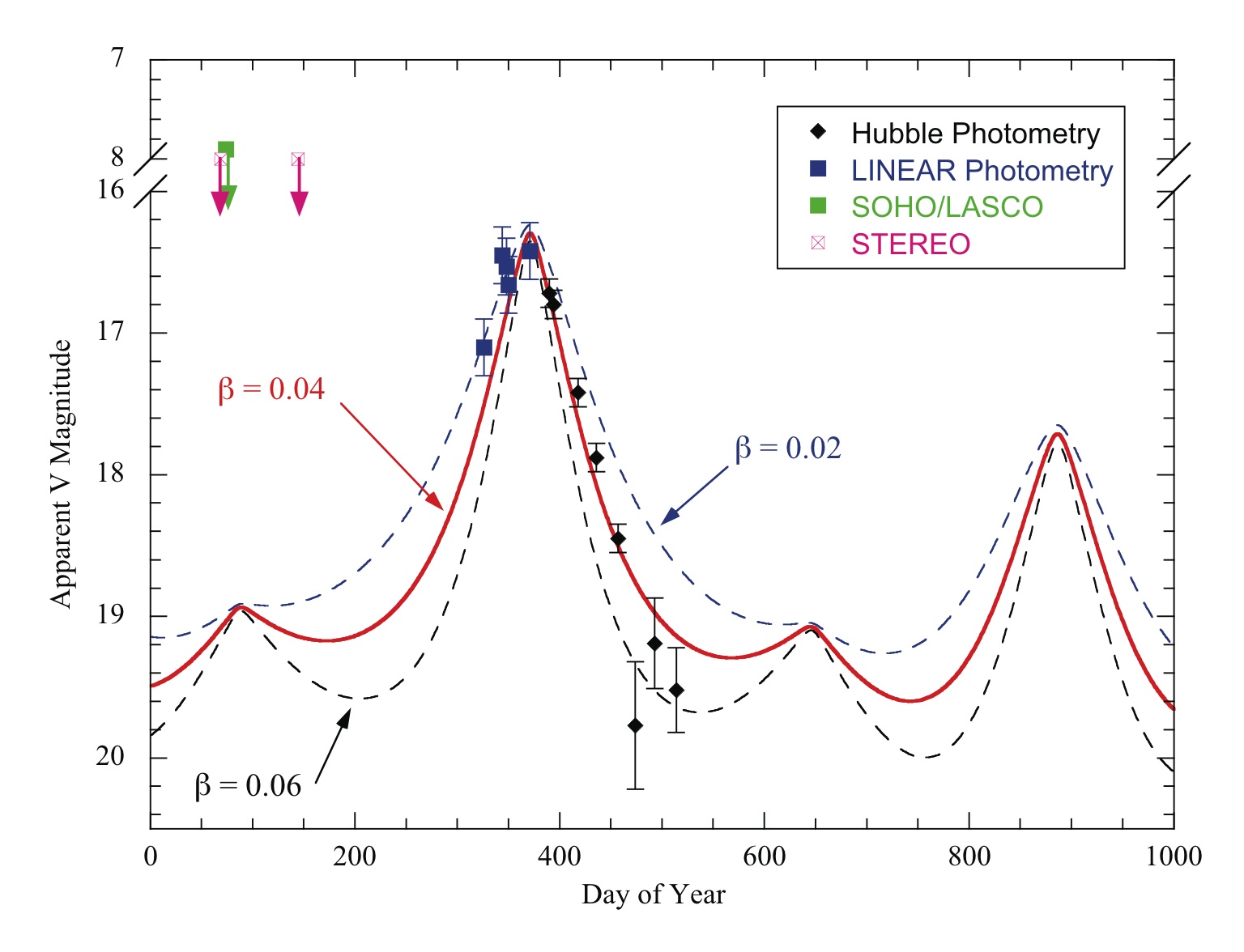}
\caption{Integrated light photometry of P/2010 A2 as a function of time, using data from the SOHO/LASCO, LINEAR and Hubble telescopes (Jewitt et al.~2010b).  Lines have the same meanings as in Figure (\ref{Dm}) but have been shifted vertically to match the photometry.  The vertical axis is broken for clarity.  \label{photometry} } 
\end{center} 
\end{figure}

Figure (\ref{photometry}) shows the available photometry of P/2010 A2, from the present work, compared with published measurements obtained from the Hubble Space Telescope (Jewitt et al.~2010b, see also Table (\ref{phot})).  Also plotted are model curves from Equation (\ref{dmag}) that have been shifted vertically to match the data by eye.   
Examination of the Figure shows that the nominal brightness of P/2010 A2 in 2009 March (DOY 74) was  $\sim$11 magnitudes fainter than the SOHO/LASCO limit.  A brightness enhancement by 2.3 to 5.5 magnitudes, as indicated by Figure (\ref{f}), would still not lift the object into detection, consistent with the fact that P/2010 A2 is not detected in the SOHO data.  To produce a brightening of 11 magnitudes would require (by Equation (\ref{factor}) and Figure (\ref{f})) a differential size index $q >$ 4.1.  

\begin{figure}
\epsscale{0.9}
\begin{center}
%%\plotfiddle{f101.ps}{0.5cm}{-90.}{.4}{.4}{-10000}{.00}
%\plotone{figure6.pdf}
%\plotone{abs_mags.pdf}
\plotone{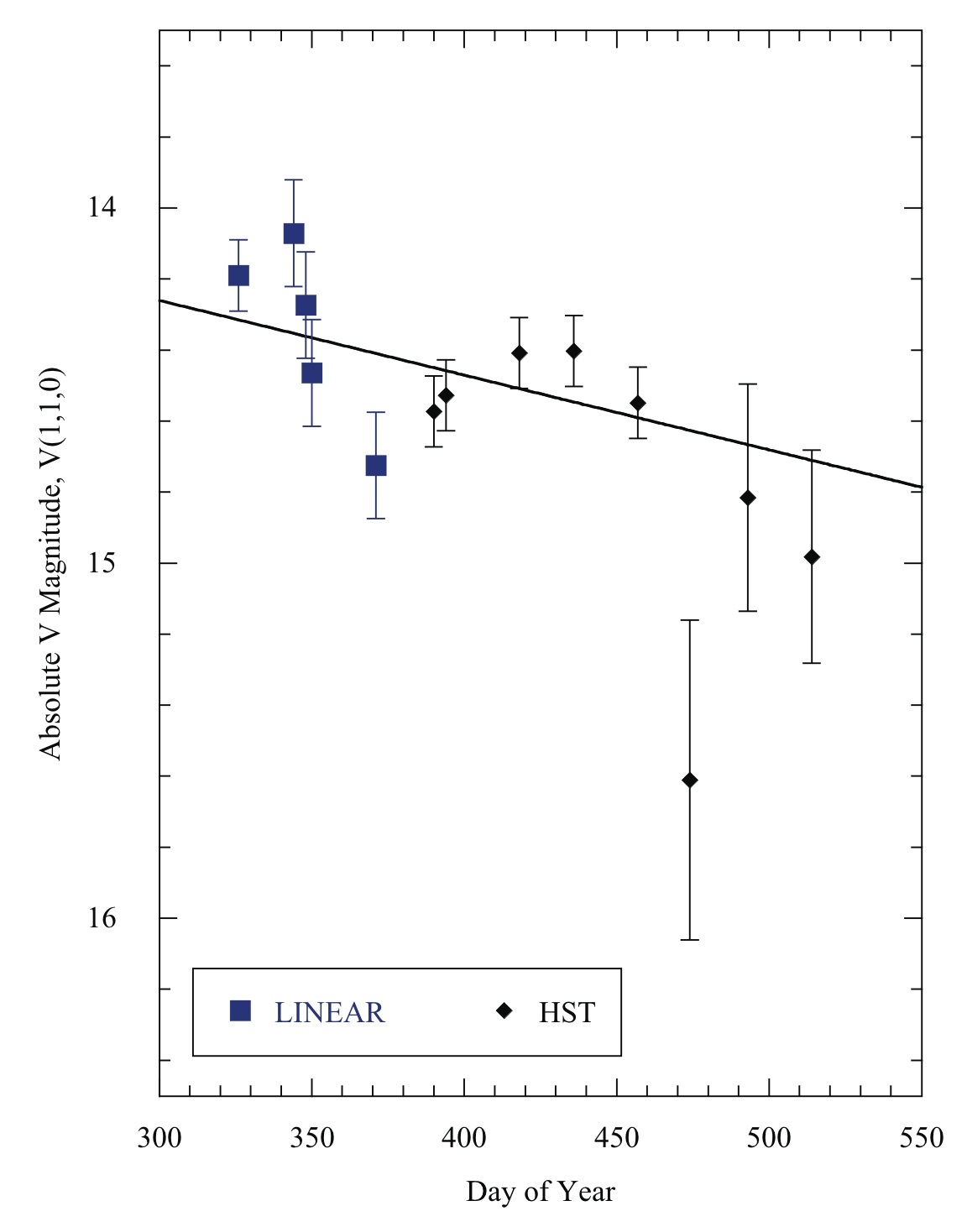}
\caption{Absolute integrated magnitudes computed from the LINEAR and HST photometry, as a function of time.  The line shows a least-squares fit to the data, weighted by the photometric uncertainties (see text).  \label{abs_mags} } 
\end{center} 
\end{figure}

The data in Figure (\ref{photometry}) show a slight asymmetry with respect to the model phase curves, falling slightly above a given curve before peak brightness and slightly below after the peak.  If real, this might indicate a fading of the object, perhaps caused by continued radiation pressure sweeping of particles from the aperture employed for photometry.  In Figure (\ref{abs_mags}) we show the absolute magnitudes, $V(1,1,0)$, computed from Equation (\ref{dmag}) with $\beta$ = 0.04 mag. degree$^{-1}$.  There is an evident trend towards larger (fainter) $V(1,1,0)$ with time.  A linear least-squares fit to $V(1,1,0) = \zeta + \eta (DOY - 371)$, weighted by the uncertainties on the measurements, gives $\zeta$ = 14.41$\pm$0.05 mag.~and $\eta$ = 0.0021$\pm$0.0009 mag.~day$^{-1}$.  This fit is shown as a solid line in Figure (\ref{abs_mags}).  The gradient in $V(1,1,0)$ lies at only the $\sim$2.3$\sigma$ confidence level, and so cannot be considered significant.  Setting $\eta$ = 0, the weighted mean absolute magnitude from the combined LINEAR and HST photometry data is $V(1,1,0)$ = 14.49$\pm$0.04 mag., which we take to be the best estimate of the absolute magnitude of P/2010 A2 in the 2009 November - 2010 May period.  The true uncertainty on $V(1,1,0)$ is probably several times larger than the formal $\pm$0.04 magnitudes, as a result of the use of non-standard bandpass photometry and other systematic effects.

The observations therefore show that the long interval between the initial brightening of P/2010 A2 in early 2009 and its eventual discovery in early 2010 has several causes.  The SOHO/LASCO and STEREO solar telescopes had sensitivity too low to detect the object against the bright background corona. Small solar elongation held the object beyond the reach of night time telescopes until about 2009 August 9 (DOY = 221, when elongation reached $\epsilon $ = 60$\degr$).   Once visible against dark sky, P/2010 A2 evaded detection by the Catalina and Spacewatch sky surveys owing to their incomplete sky coverage.  The object was too faint to be detected by LINEAR until UT 2009 Nov 22, but was missed in LINEAR data by automatic object-finding software until UT 2010 Jan 06, as a result of the low surface brightness of the tail and of interference by overlapping background stars.  Detection before Nov 22 was thwarted, in addition, by geometrical dimming caused by a larger geocentric distance.  Furthermore, the apparent brightness of P/2010 A2 faded by a magnitude or more in the month following discovery (Figure \ref{photometry}), showing the limited opportunities for detection from the LINEAR telescope.

Figure (\ref{photometry}) shows that LINEAR could only detect P/2010 A2 when $V \leq$ 17.2 (and even then, perhaps not with 100\% efficiency, given the deleterious effects of field star contamination).  We determined that an object with the absolute magnitude (assumed constant) and orbit of P/2010 A2 will appear at $V \leq$ 17.2 for only $\sim$6\% of the 20 year period from 2000 to 2019.   Thus, we may estimate the probability of detection of P/2010 A2 clones in LINEAR data as $\leq$6\%, even if they do not fade as a result of radiation pressure sweeping of dust.  Intrinsically fainter, or more distant, objects would have an even smaller probability of detection. Still, for every example detected, another $\sim$15 or more go unnoticed because they are projected in the daytime sky, or are too distant and faint or fall beneath the detection threshold for other reasons.  Disrupting asteroids have not been previously observed mainly because they are challenging observational targets.  More positively, this first detection, made apparently against the observational odds, raises the prospect that a substantial population of recently-disrupted asteroid remnants lies awaiting detection by more sensitive surveys.  The accumulation of a sample of similar objects will allow us to examine impulsive dust production in real-time, so opening a new window onto the study of asteroid impact.  Recently reported activity in (596) Scheila (Larson~2010) already provides a second example of the immediate aftermath of a recent impact (Bodewits et al.~2011, Jewitt et al.~2011).

\clearpage

\section{Summary}
We searched for pre-discovery observations of recently disrupted asteroid P/2010 A2 to try to understand the long interval between its disruption (in 2009 Feb/Mar) and discovery (2010 Jan 06), with the following results:

\begin{enumerate}
\item The disruption event in early 2009 occurred at a small solar elongation ($\le$ 30$\degr$). The elongation remained $<$60$\degr$ as seen from Earth until about 2009 August,  explaining the non-detection of P/2010 A2 by night-time telescopes in this early period.  

\item Observations in the daytime sky using the SOHO and STEREO solar satellites  placed only upper limits to the allowable brightness.  The non-detections require that the differential size distribution index, extended down to 0.1 $\mu$m particle sizes, should be $q <$ 4.1. This is consistent with the value $q$ = 3.3$\pm$0.2 measured in the $mm$ to $cm$ size range (Jewitt et al.~2010b).

%\item The geometric dimming reached a minimum on UT 2010 Jan 6, coincident with the discovery by LINEAR on the same day.   

\item  The date of eventual discovery, UT 2010 Jan 06, coincides exactly with the date of peak apparent brightness as seen from Earth, showing the importance of observational selection effects in the discovery.   

\item Pre-discovery observations of P/2010 A2 by the LINEAR survey telescope were identified on UT 2009 Nov 22 and Dec 10, 15 and 16.  They and later observations show an object whose integrated brightness varies primarily in response to the changing observing geometry, meaning that the total scattering cross-section remained approximately constant after 2009 November.  The absolute magnitude and its formal uncertainty are $V(1,1,0)$ = 14.49$\pm$0.04 mag.

\item The fraction of the time spent by P/2010 A2  above the LINEAR detection threshold is $\leq$ 6\%, while intrinsically fainter analogs of this object would be even less likely to be detected.  We conclude that similar examples of recently impacted asteroids are common, and will be revealed by future all-sky surveys in large numbers.

\end{enumerate}

\acknowledgements
We thank Ed Beshore and Robert McMillan for checking the data archives of the Catalina and Spacewatch sky surveys, respectively, at our request and Michal Drahus, Henry Hsieh and an anonymous referee for comments on the manuscript.  The LINEAR program is sponsored by the National Aeronautics and Space Administration (NRA No.
NNH06ZDA001N, 06-NEOO06-0001) and the United States Air Force under Air Force Contract
FA8721-05-C-0002. Opinions, interpretations, conclusions, and recommendations are those of the
authors and are not necessarily endorsed by the United States Government.  This work was supported in part by a NASA grant to DJ.

\clearpage

\begin{deluxetable}{lllllcr}
%\tabletypesize{\scriptsize}
\tablecaption{Photometry
\label{phot}}
\tablewidth{0pt}
\tablehead{
\colhead{Observatory} &\colhead{UT\tablenotemark{a}} & \colhead{DOY\tablenotemark{b}} & \colhead{$V$  \tablenotemark{c}} & \colhead{$R$ [AU]\tablenotemark{d}} & \colhead{$\Delta$ [AU]\tablenotemark{e}}   & \colhead{$\alpha$ [$\degr$]\tablenotemark{f}} }
\startdata
%30  $\le \alpha <$ 40 &	0.172 & 	1.094 & 19 & 11.35$\pm$0.19 \\
%year  &  doy  &  integ   & skyMag  &  limMag  &  detMag   \\
STEREO A & 2009 Feb 10 & 41 & $>$8.0 & 2.299 & 3.242 & 3.2 \\
SOHO & 2009 Mar 15 &    74      & $>$7.9   &   2.252   &    3.226   & 3.0 \\
STEREO B & 2009 May 24 & 144  & $>$8.0 & 2.154 & 3.171 & 3.4 \\
LINEAR & 2009  Nov 22 &   326    &    17.10$\pm$0.10   &    2.006   &    1.246   &   23.0      \\     
LINEAR & 2009  Dec 10 &   344   &     16.45$\pm$0.15  &     2.006   &    1.112  &    15.9     \\      
LINEAR & 2009  Dec 15  &   349   &    16.53$\pm$0.15   &    2.006   &    1.084   &    13.5     \\      
LINEAR & 2009  Dec 16 &   350   &     16.66$\pm$0.15   &    2.006   &    1.079   &    12.9      \\   
LINEAR & 2010 Jan 06 & 371 & 16.42$\pm$0.15 & 2.010 & 1.030 & 2.9 \\
HST & 2010 Jan 25 & 390 & 16.72$\pm$0.08 & 2.018 & 1.078 & 11.5 \\
HST & 2010 Jan 29 & 394 & 16.80$\pm$0.08 & 2.020 & 1.099 & 13.5 \\
HST & 2010 Feb 22 & 418 & 17.42$\pm$0.08 & 2.034 & 1.286 & 23.1 \\
HST & 2010 Mar 12 & 436 & 17.88$\pm$0.08 & 2.047 & 1.473 & 27.0 \\
HST & 2010 Apr 02 & 457 & 18.45$\pm$0.08 & 2.066 & 1.717 & 28.8 \\
HST & 2010 Apr 19 & 474 & 19.77$\pm$0.45 & 2.084 & 1.922 & 28.7 \\
HST & 2010 May 08 & 493 & 19.19$\pm$0.32 & 2.105 & 2.150 & 27.4 \\
HST & 2010 May 29 & 514 & 19.52$\pm$0.30 & 2.130 & 2.393 & 25.0 \\
\enddata

%% Text for table notes should follow after the \enddata but before
%% the \end{deluxetable}. Make sure there is at least one \tablenotemark
%% in the table for each \tablenotetext.

\tablenotetext{a}{UT Date of the observation}
\tablenotetext{b}{Day of Year (DOY = 1 on UT 2009 Jan 01)}
\tablenotetext{c}{Apparent $V$ magnitude}
\tablenotetext{d}{Heliocentric distance}
\tablenotetext{e}{Object to observatory distance}
\tablenotetext{f}{Phase angle}

\end{deluxetable}

\clearpage

\clearpage

\clearpage

\clearpage

\clearpage

%\clearpage

%\begin{figure}
%\epsscale{1.0}
%\begin{center}
%%%\plotfiddle{f101.ps}{0.5cm}{-90.}{.4}{.4}{-10000}{.00}
%%\plotone{figure6.pdf}
%\plotone{elong_vs_date.pdf}
%\caption{Elongation of P/2020 A2 vs. date (expressed as Julian Day Number), as seen from the SOHO and STEREO A and B observatories.  The height of each colored box shows the range of elongations observable from the respective observatory, while the width of each box shows the range of dates within which the elongation  fell within the field of view.   \label{elong} } 
%\end{center} 
%\end{figure}

\clearpage

\clearpage

\clearpage

\end{document}